\documentstyle[twoside,fleqn,espcrc2]{article}

\def\lsim{\ ^<\llap{$_\sim$}\ }
\def\gsim{\ ^>\llap{$_\sim$}\ }
\def\r2{\sqrt 2}
\def\rmuu{\gamma^{\mu}}

\def\PL{{1-\gamma_5\over 2}}
\def\PR{{1+\gamma_5\over 2}}
\def\T3f{T_{3f}}
\def\sinW2{\sin^2\theta_W}

\def\v#1{v_#1}
\def\tb{\tan\beta}
\def\c2b{\cos 2\beta}

\def\KK{$K^0$-$\bar{K^0}$}
\def\wi{\omega_i}
\def\xj{\chi_j}
\def\sf{\tilde f} 
\def\Wmu{W_\mu}

\def\mgr{m_{3/2}}
\def\m#1{{\tilde m}_#1}
\def\mH{m_H}
\def\mw#1{{\tilde m}_{\omega #1}}
\def\mx#1{{\tilde m}_{\chi #1}}

\def\MsfL{{\tilde M}_{fL}} 
\def\MsfR{{\tilde M}_{fR}} 
\def\Msfl{{\tilde M}_{f1}} 
\def\Msfh{{\tilde M}_{f2}}

\def\Fq{F(q^2)}

\newcommand{\AmS}{{\protect\the\textfont2
  A\kern-.1667em\lower.5ex\hbox{M}\kern-.125emS}}

\title{Electric dipole moments of W boson and neutron 
          from supersymmetry}

\author{Noriyuki Oshimo\address{Department of Physics, 
                 Ochanomizu University,  \\ 
           Otsuka 2-1-1, Bunkyo-ku, Tokyo 112, Japan}%
           } 
\begin{document}

\begin{abstract}
     The supersymmetric standard model (SSM) has a CP-violating phase  
in its gauge-Higgs sector.  This new source of CP violation  
induces the electric dipole moments (EDMs) 
of the neutron and electron at the one-loop level.    
The same CP-violating phase generates the EDM of the $W$ boson  
through one-loop diagrams mediated by the charginos and neutralinos,  
which leads to 
the EDMs of the neutron and electron at the two-loop level.  
We discuss these EDMs, assuming that the CP-violating phase  
has the magnitude 
of order 0.1--1 which may be suggested from the baryon 
asymmetry of the universe.     
Even at the two-loop level the neutron and the electron  
could have EDMs which are smaller than their experimental 
upper bounds by only one order of magnitude.  
Since these two-loop contributions 
do not depend on the values of SSM parameters
for the squark or slepton sector, they provide predictions for the 
EDMs less ambiguous than the one-loop contributions.   

\end{abstract}

\maketitle

\section{Introduction}

     Nature is not invariant under CP transformation 
as observed in the \KK\ system \cite{CPrev}.  In the 
standard model (SM)  
CP violation originates from the Kobayashi-Maskawa (KM) 
phase, which is consistent with all the experimental 
results ever found concerning CP violation.  
On the other hand, it has been suggested that 
the baryon asymmetry of the universe 
could also be an outcome of CP violation at the electroweak
scale \cite{BArev}.  
However, this asymmetry has been shown not to be  
explained quantitatively within the framework of the SM.  
If the baryon asymmetry was really generated at the 
electroweak phase transition of the universe, some extension 
has to be introduced for the SM.  

     The supersymmetric standard model (SSM) is one 
of the most plausible extensions of the SM, 
which has new sources of CP violation \cite{ellis} in addition 
to the KM phase.  In this model 
the CP-violating phenomena in the \KK\ system  
are described by the KM phase, while the baryon 
asymmetry of the universe could come from a new 
CP-violating phase which is contained in the 
gauge-Higgs sector.  Indeed, it has been shown \cite{BA} 
that the CP-violating interactions for the charginos 
generate the asymmetry and 
the resultant ratio of baryon number to entropy 
becomes consistent with its observed value, if the 
CP-violating phase is of order $0.1-1$.  
In this case the new source of CP violation 
can sizably affect the other phenomena.  
It would be important to study 
the effects of the CP-violating phase on 
experimentally measurable quantities.  

      In this report we discuss the 
electric dipole moments (EDMs)  
of the neutron and electron together with 
that of the $W$ boson within the framework of the SSM.  
Owing to the new CP-violating phase 
in the gauge-Higgs sector,  
the neutron and electron EDMs receive contributions from  
diagrams mediated by the charginos and squarks or sleptons 
at the one-loop level \cite{ellis,edm,edmnew}.   
In addition, the same CP-violating phase 
yields the EDM of the $W$ boson through one-loop 
diagrams mediated by the charginos and neutralinos. 
The $W$-boson EDM can then  
induce the neutron and electron EDMs through one-loop 
diagrams generated by the SM interactions \cite{1wedm,wedm}.  
Taking into account these two-types of contributions,   
we make detailed analyses of the EDMs.  

     The EDMs of the neutron and electron from the 
$W$-boson EDM are generated at the two-loop level.  
For the CP-violating phase of order unity these EDMs 
have values smaller than their present experimental 
upper bounds by one order of magnitude \cite{wedm}.  
The two-loop contributions are a priori expected to be smaller 
than the contributions from the one-loop diagrams.  
However, the two-loop contributions 
from the $W$-boson EDM are determined 
only by the gauge-Higgs sector, while the one-loop contributions 
are determined not only by the gauge-Higgs sector 
but also by the squark or slepton sector, 
especially by their masses.    
Consequently, as the squark or slepton masses  
increase, the one-loop contributions become small, 
while the two-loop contributions are kept the same.  
The values of the neutron and  
electron EDMs induced by the $W$-boson EDM 
are predicted with less uncertainty in the SSM.  

     The present experimental bounds for  
the EDMs of the neutron and electron are given by  
$|d_n|\lsim 10^{-25}e$cm \cite{nEDM} 
and $|d_e|\lsim 10^{-26}e$cm \cite{eEDM}, respectively.  
In the SM, the EDM of the neutron vanishes 
at both the one-loop and the two-loop
levels, resulting in $|d_n|<10^{-30}e$cm, 
and the EDM of the electron is much smaller \cite{CPrev}.    
Therefore, the detection of these EDMs in the foreseeable future   
may be considered to  
provide an indirect evidence for 
the existence of supersymmetry in nature.  

\section{CP-violating interactions}

     The SSM is an extension of the SM based on  
$N=1$ supergravity coupled to grand unified 
theories (GUTs) \cite{SUSYrev}.  
This model contains several parameters 
whose values are generally complex.    
Although there is some freedom for the phases 
of particle fields, the redefinitions of the fields 
cannot rotate away all the complex phases.  
Even if generation mixings among matter fields are neglected, 
at least two of the complex parameters cannot be made real.   
One physical complex phase appears in the gauge-Higgs sector 
and another in the squark and slepton sector.  
These are the new sources of CP violation intrinsic in the SSM.  
In this section we briefly review the CP-violating interactions 
relevant to our discussions.  

     The charginos $\wi$ and the neutralinos $\xj$  
are charged and neutral mass eigenstates 
for the gauginos and Higgsinos, the superpartners of the 
gauge and Higgs bosons.  Their mass matrices are given by 
\begin{equation}
    M^- = \left(\matrix{\m2 & -g\v1^*/\r2 \cr
                -g\v2^*/\r2 & \mH}        \right) 
\label{1} 
\end{equation}
for the charginos, and 
\begin{eqnarray}
\lefteqn{M^0 = }          \nonumber \\ 
\lefteqn{\left(\matrix{\m1 &  0  & g'\v1^*/2 & -g'\v2^*/2 \cr
                         0  & \m2 & -g\v1^*/2 &   g\v2^*/2 \cr
                       g'\v1^*/2 & -g\v1^*/2 &   0  & -\mH \cr
                      -g'\v2^*/2 &  g\v2^*/2 & -\mH &   0}
           \right)  }
\label{2}
\end{eqnarray}
for the neutralinos.  
Here $\v1$ and $\v2$ denote  
the vacuum expectation values of the  
Higgs bosons with U(1) hypercharges $-1/2$ and $1/2$, respectively;   
$\m2$ and $\m1$ the SU(2) and U(1) gaugino masses,  
respectively, which appear in supersymmetry soft-breaking terms
of the Lagrangian;  
and $\mH$ the mass parameter in a bilinear term of 
Higgs superfields in superpotential.    
The mass matrices (\ref{1}) and (\ref{2}) are diagonalized by unitary 
matrices $C_R$, $C_L$, and $N$ as
\begin{equation}
      C_R^\dagger M^-C_L = {\rm diag}(\mw1, \mw2) \quad 
                       (\mw1 <\mw2 )   
\label{3}
\end{equation}
and
\begin{eqnarray}
N^tM^0N &=& {\rm diag}(\mx1, \mx2, \mx3, \mx4) 
\label{4} \\
                & &       (\mx1<\mx2<\mx3<\mx4), \nonumber  
\end{eqnarray}
giving the mass eigenstates.  

     The squarks and sleptons are superpartners of the 
quarks and leptons.  For each flavor there are two mass 
eigenstates $\sf_i$ which are composed of $\sf_L$ and $\sf_R$, 
the scalar partners for the left-handed and right-handed 
components of the fermion $f$.  
Their mass-squared matrix becomes
\begin{equation}
M^2_f = \left(\matrix{(M^2_f)_{11} & (M^2_f)_{12} \cr 
                        (M^2_f)_{21} & (M^2_f)_{22} }
           \right), 
\label{5} \\
\end{equation}
\begin{eqnarray}
\lefteqn{(M^2_f)_{11}=m_f^2 + \c2b (\T3f - Q_f\sinW2 )M_Z^2 } 
		    \nonumber \\
\lefteqn{    + \MsfL^2, } \nonumber \\
\lefteqn{(M^2_f)_{12}=m_f (R_f\mH + A_f^*\mgr), } \nonumber \\
\lefteqn{(M^2_f)_{21}=m_f (R_f^*\mH^* + A_f\mgr), } \nonumber \\
\lefteqn{(M^2_f)_{22}=m_f^2 + Q_f\c2b\sinW2 M_Z^2 + \MsfR^2, } 
		  \nonumber 
\end{eqnarray}
where $R_f$ and $\tb$ are defined by 
\[
R_f = \left\{\begin{array}{ll}
             \v1/\v2^* &  
         \mbox {for}\ \T3f=1/2 \\
                    \v2/\v1^* & 
         \mbox {for}\ \T3f=-1/2   
        \end{array}\right.  ,
\] 
\begin{equation}
\tb = |\v2 /\v1|. 
\label{6}
\end{equation}
Here $m_f$ represents a mass of the fermion $f$;
$Q_f$ an electric charge; 
$\T3f$ the third component of the weak isospin of the left-handed 
component of $f$;
$A_f$ a dimensionless constant  
which originates in a trilinear coupling of the 
squarks or sleptons in supersymmetry soft-breaking terms;   
$\mgr$ the gravitino mass; and $\MsfL^2$ and $\MsfR^2$ 
mass-squared parameters for $\sf_L$ and $\sf_R$, respectively.
We have neglected generation mixings.  
The mass-squared matrix (\ref{5}) 
is diagonalized by a unitary matrix $S_f$ as
\begin{equation}
      S_f^{\dag}M^2_f S_f = {\rm diag}(\Msfl^2, \Msfh^2),
\label{7}
\end{equation}
giving the mass eigenstates.  
 
     The SSM parameters 
appearing in eqs. (\ref{1}) and (\ref{2})   
are $\v1$,
$\v2$, $\m2$, $\m1$, and $\mH$.  
We assume the GUT relation 
$\m1=(5/3)\tan^2\theta_W\m2$.  The redefinitions of 
the fields make it possible without loss of 
generality to take all these parameters 
except $\mH$ real and positive.  Then, 
the CP-violating phase is represented by the phase of $\mH$, 
which we express as  
\begin{equation}
\mH=|\mH|\exp(i\theta).  
\label{8}
\end{equation}
Since $\v1$ and $\v2$ are related to the $W$-boson mass $M_W$, 
the independent parameters become $\tb$, $\m2$, $|\mH|$, 
and $\theta$. 

     The mass-squared matrix (\ref{5}) 
depends on various SSM parameters.  
The parameter $A_f$ has a physical complex phase different from $\theta$.  
For the squarks or sleptons of the first generation, however, 
the matrix is approximately proportional to the unit matrix 
because of the smallness of the corresponding quark 
or lepton mass.  In this approximation the matrix is determined 
by a mass-squared parameter for the squark or slepton masses.  

\begin{figure}[htb]
\vspace{8cm}
\includegraphics{edm1.ps}
\caption{The Feynman diagrams for the EDM of a quark or a lepton.}
\label{fig1}
\end{figure}
     The complex mass matrices for the charginos and 
neutralinos lead 
to CP-violating interactions of the charginos, neutralinos, 
and $W$ bosons.  
The interaction Lagrangian for these particles is given by 
\begin{eqnarray}
\lefteqn{{\cal L} =}   \nonumber \\
\lefteqn{      \frac{1}{\r2}g\bar{\xj}\rmuu
                  \left(G_{Lji}\PL+G_{Rji}\PR\right)
                           \wi\Wmu^\dagger }\nonumber \\
\lefteqn{                  +{\rm H.c.},  }  
\label{9}    \\
\lefteqn{ G_{Lji} = \r2 N^*_{2j}C_{L1i}+N^*_{3j}C_{L2i}, } \nonumber  \\
\lefteqn{ G_{Rji} = \r2 N_{2j}C_{R1i}-N_{4j}C_{R2i}.  }\nonumber  
\end{eqnarray}   
Since the coupling constants $G_{Lji}$ and $G_{Rji}$ have 
different 
complex phases, this Lagrangian is not invariant under 
CP transformation.  

     The interaction Lagrangian for the charginos, 
$u$-type quark, and 
$d$-type squarks and that for the charginos, $d$-type quark, and 
$u$-type squarks are respectively given by 
\begin{eqnarray}
\lefteqn{{\cal L}=}  \nonumber \\
\lefteqn{ig\bar{\wi^c} \left(A_{Li}^k\PL+A_{Ri}^k\PR\right)
                           u{\tilde d}_k^\dagger   +{\rm H.c.},}    
\label{10}    \\
\lefteqn{A_{Li}^k = C_{L1i}S_{d1k}^*
                      -\frac{\eta_d}{g}C_{L2i}S_{d2k}^*,}  \nonumber \\
\lefteqn{A_{Ri}^k = -\frac{\eta_u^*}{g}C_{R2i}S_{d1k}^*,}    \nonumber  
\end{eqnarray}
and
\begin{eqnarray}
\lefteqn{{\cal L}=} \nonumber \\
\lefteqn{ig\bar{\wi} \left(B_{Li}^k\PL+B_{Ri}^k\PR\right)
                           d{\tilde u}_k^\dagger   +{\rm H.c},}
\label{11}   \\
\lefteqn{B_{Li}^k = C_{R1i}^*S_{u1k}^*
                     +\frac{\eta_u}{g}C_{R2i}^*S_{u2k}^*,} \nonumber \\ 
\lefteqn{B_{Ri}^k = \frac{\eta_d^*}{g}C_{L2i}^*S_{u1k}^*.}   \nonumber  
\end{eqnarray}
Here $\eta_f$ $(f=u, d)$ represents 
the Yukawa coupling constant 
for the $f$-type quark.  
The matrices $S_u$ and $S_d$ are 
approximately the unit matrix for the $u$- and $d$-squarks.    
The interaction Lagrangians for the charginos, 
leptons, and sleptons are 
obtained by appropriately changing eqs. (\ref{10}) and (\ref{11}). 

\section{EDM of neutron at one-loop level}

\begin{figure}[htb]
\vspace{6cm}
\includegraphics{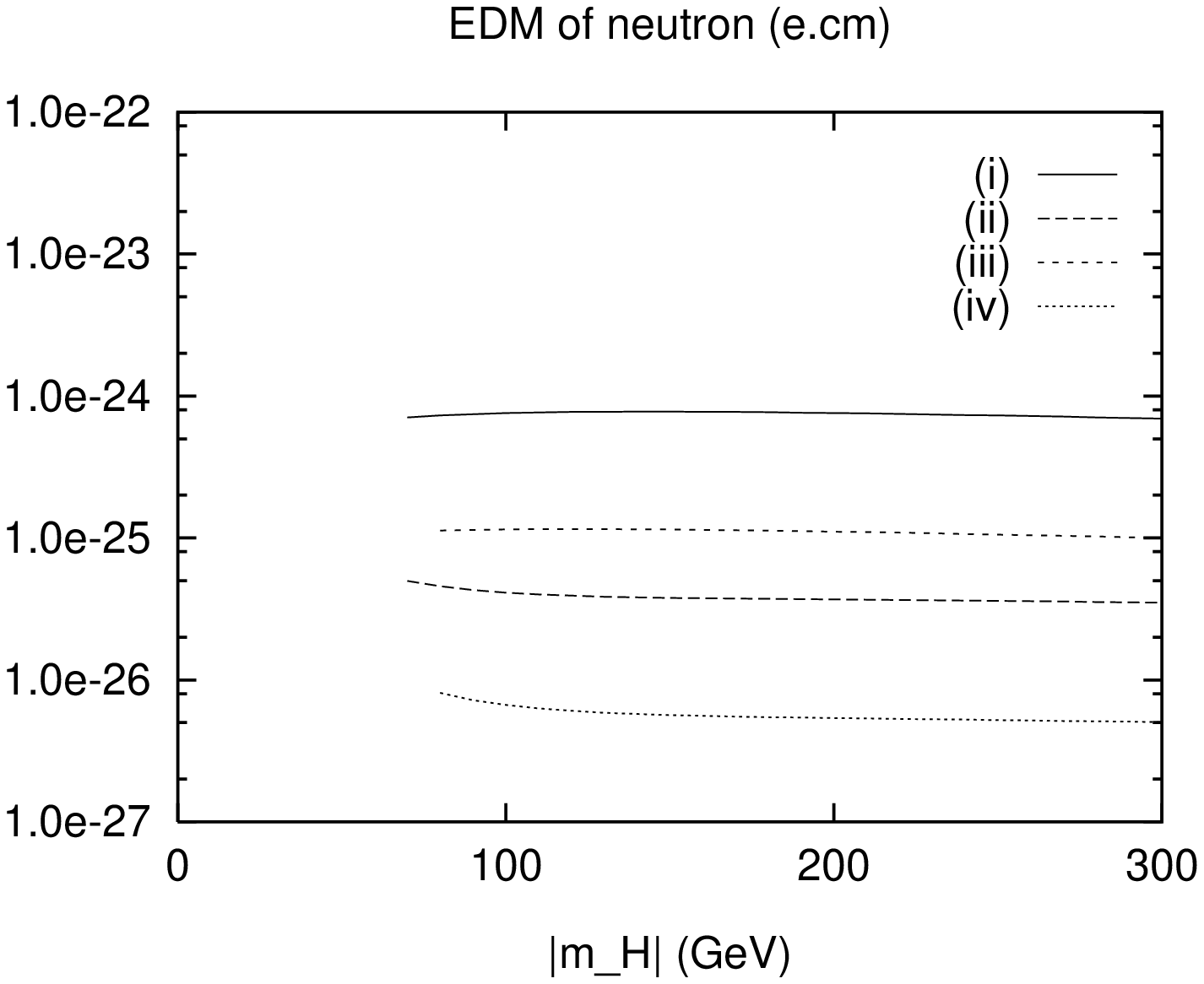}
\caption{The EDM of the neutron at the one-loop level for 
       $\tb=2$ and $\m2=200$ GeV.  
        The values of $\theta$ and the squark mass for 
	curves (i)--(iv) are listed in Table 1.}
\label{fig2}
\end{figure}         
\begin{table}[hbt]
\caption{The values of $\theta$ and the squark mass for curves 
            (i)--(iv) in Fig. 2.}
\label{tab1}
\begin{tabular}{lcccc}
\hline
   & (i) & (ii) & (iii) &(iv) \\
\hline
$\theta$  & $\pi/4$  & $\pi/4$  & $0.1$ & $0.1$  \\
Squark mass (TeV) & $1$ & $5$ & $1$ & $5$ \\
\hline
\end{tabular}
\end{table}
     In the SSM the EDM of the quark   
receives contributions from one-loop diagrams in which 
the charginos, neutralinos, or gluinos are exchanged together 
with the squarks.  
Assuming the nonrelativistic quark model, 
the EDM of the neutron $d_n$ is given, in terms of 
the $u$-quark EDM $d_u$ and the $d$-quark EDM $d_d$, 
by $d_n=(4d_d-d_u)/3$.  
The electron EDM is also induced by one-loop diagrams, 
where the charginos or neutralinos are exchanged with the sleptons.  
Among these diagrams, the chargino-mediated 
ones generally make dominant contributions 
for both the neutron and the electron EDMs \cite{edm}.  
The relevant Feynman diagrams are depicted in Fig. \ref{fig1}.

\begin{figure}[htb]
\vspace{6cm}
\includegraphics{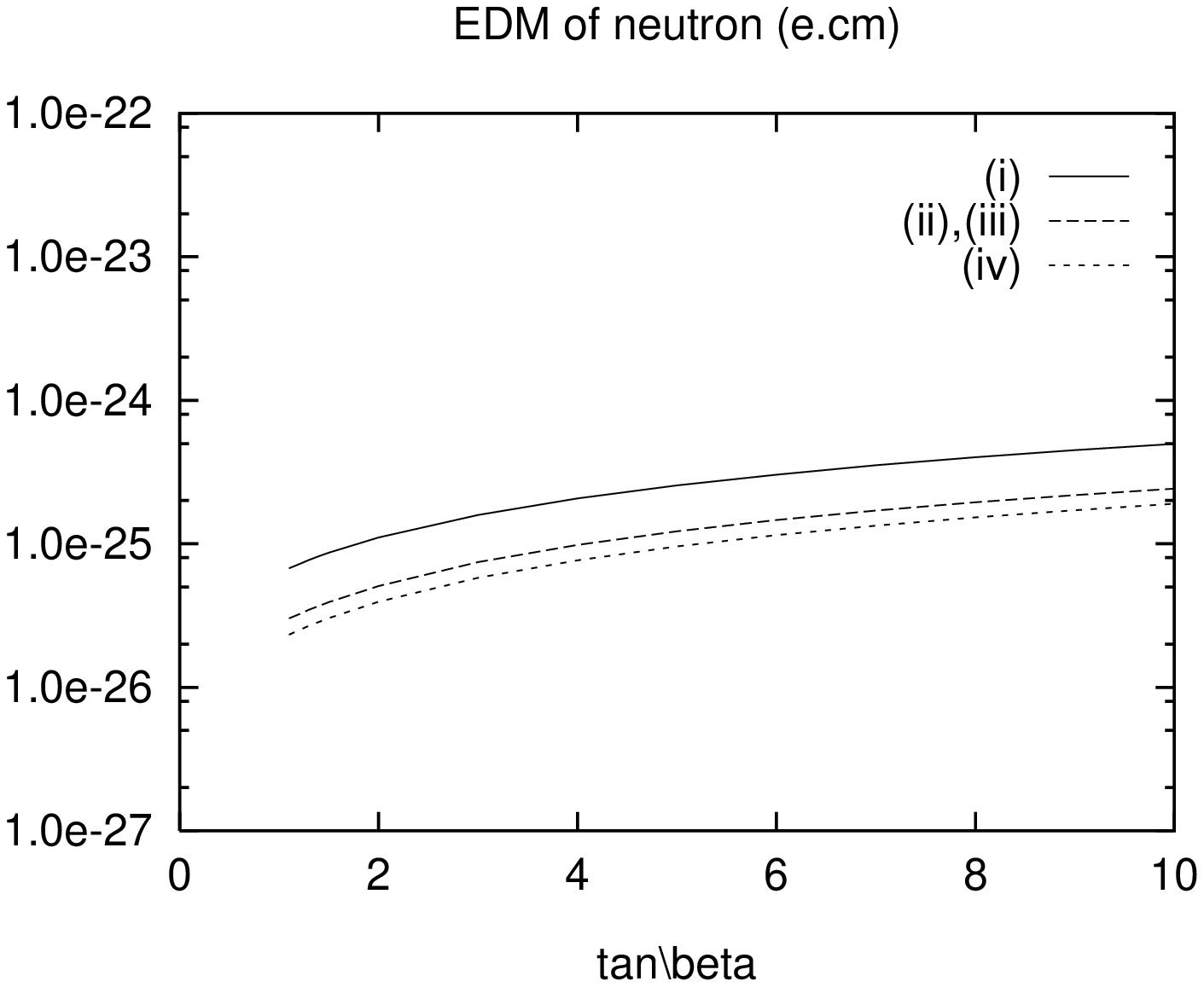}
\caption{The EDM of the neutron at the one-loop level  
        for $\theta=0.1$.  
	The squark mass is taken for 1 TeV and 
        the values of $\m2$ and $|\mH|$ for curves (i)--(iv) 
         are listed in Table 2.}
\label{fig3}
\end{figure}         
\begin{table}[hbt]
\caption{The values of $\m2$ and $|\mH|$ for curves 
                (i)--(iv) in Figs. 3 and 6.}
\label{tab2}
\begin{tabular}{lcccc}
\hline
      & (i) & (ii) & (iii) &(iv) \\
\hline
 $\m2$ (GeV) & 200 & 200 & 1000 & 1000 \\
$|\mH|$  (GeV) & 200  & 1000  & 200 & 1000  \\
\hline
\end{tabular}
\end{table}
     We show in Fig. \ref{fig2} the absolute value of 
the neutron EDM for $\tb=2$ and $\m2=200$ GeV 
as a function of $|\mH|$. 
For $\theta$ and the squark mass we have taken four sets of values given in 
Table \ref{tab1}.   
In the ranges of $|\mH|$ where curves are not drawn, 
the lighter chargino has a mass smaller than 45 GeV which has been 
ruled out by LEP experiments \cite{PDG}.  
The experimental constraints on the neutron EDM are 
satisfied, if the masses of the  squarks are larger than or around 1 TeV 
for $\theta\gsim 0.1$, 
whereas the masses of the charginos and  
neutralinos can be of order of 100 GeV \cite{edm}.  
The analyses of the EDM of the electron give approximately the same 
constraints on the masses of the sleptons, charginos, and 
neutralinos.  

     In Fig. \ref{fig3} the neutron EDM is plotted as a 
function of $\tb$ for four sets of values of $\m2$ and $|\mH|$ 
given in Table \ref{tab2}.  
The CP-violating phase $\theta$ and the squark mass are taken 
for 0.1 and 1 TeV, respectively.  
Curves are not drawn for $\tb<1$, because 
the value of $\tb$ is theoretically considered not to be smaller than 
unity, if the SU(2)$\times$U(1) gauge symmetry is broken through 
radiative corrections.  
The EDM of the neutron increases as $\tb$ increases, which also holds 
for the EDM of the electron.   
For $\tb\gsim 10$ the experimental results give constraints 
much severer than for $\tb\sim 1$.  
It should be noted that in the SSM the radiative $b$-quark decay 
$b\rightarrow s\gamma$ also receives contributions 
dominantly from one-loop diagrams mediated by the 
charginos and squarks, and has similar dependence on $\tb$ \cite{bsg}.  

\section{EDM of neutron at two-loop level}

\begin{figure}[htb]
\vspace{6cm}
\includegraphics{wedm3.ps}
\caption{The Feynman diagram for the EDM of a quark or a lepton 
              which involves a CP-violating coupling for the 
              $W$-bosons and photon.}
\label{fig4}
\end{figure}         
    The interactions in eq. (\ref{9}) give rise to 
a CP-violating anomalous coupling of the $W$ bosons and  
photon through one-loop diagrams 
in which the charginos and neutralinos are exchanged.  
The induced CP-violating term in the effective Lagrangian 
for the $W$ boson on mass-shell is   
written as  
\begin{eqnarray}
\lefteqn{{\cal L}_{eff} =} \nonumber \\
\lefteqn{ie\Fq W_\mu^\dagger W_\nu\frac{1}{2}
\epsilon^{\mu\nu\rho\sigma}
(\partial_\rho A_\sigma -\partial_\sigma A_\rho),}         
\label{12} 
\end{eqnarray}
where $\Fq$ represents a form factor, 
$q^2$ being the momentum-squared of the photon.  
The EDM of the $W$ boson is given by 
\begin{equation}
   d_W=-\frac{e}{2M_W}F(0).
\label{13}
\end{equation}
For the $W$ bosons and a vector boson, in general,  
there can be two more couplings which break CP invariance \cite{hagi}.  
However, the interactions in eq. (\ref{9}) 
only contribute to the coupling of eq. (\ref{12}).  

     The SSM parameters which determine $\Fq$ are those 
contained in eqs. (\ref{1}) and (\ref{2}).  
For $\tb\sim 1$, $\theta\sim 1$, and $\m2,|\mH|\sim 100$ GeV
the magnitude of $\Fq$ is around $10^{-4}$ \cite{wedm}, 
which is far larger than the prediction of the SM.  
The $W$-boson EDM could in principle be measured in 
$e^+e^-$ colliding experiments \cite{schild}, 
although such a magnitude for $\Fq$ would still be too small 
for detection at LEP II.  
 
     The EDM of the $W$-boson yields the EDMs of the quarks and 
leptons through one-loop diagrams.  In the SSM, therefore, 
the EDMs of the quarks and leptons 
can be induced at the two-loop level.  
The relevant diagram is shown in Fig. \ref{fig4},  
where $f$ and $f'$ denote the quarks or leptons whose  
left-handed components form an SU(2) doublet.   
At the two-loop level there also exist other diagrams 
which involve squarks or 
sleptons and make contributions to the quark or lepton EDM.  
However,
as long as the squarks and sleptons are much heavier than 
the charginos and neutralinos, these diagrams can be safely 
neglected.  
This may be indeed the case for $\theta\gsim 0.1$,  
since the squarks and sleptons should be heavier than or 
around 1 TeV from the analyses of the one-loop contributions 
to the EDMs as discussed in the previous section.  

\begin{figure}[htb]
\vspace{6cm}
\includegraphics{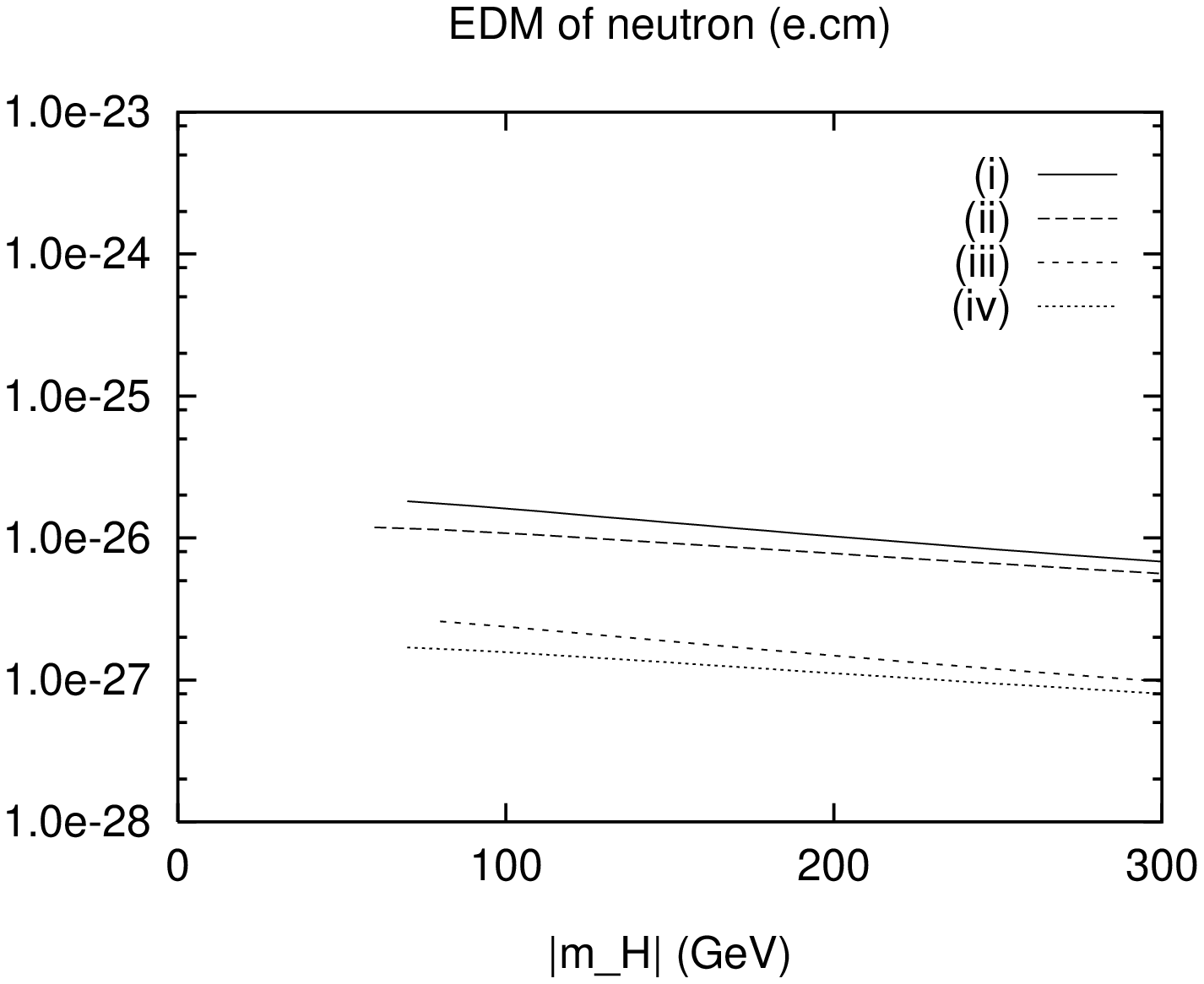}
\caption{The EDM of the neutron induced by the 
        $W$-boson EDM for $\tb=2$.  
        The values of $\theta$ and $\m2$ for curves (i)--(iv) 
         are listed in Table 3.}
\label{fig5}
\end{figure}         
\begin{table}[hbt]
\caption{The values of $\theta$ and $\m2$ for curves 
            (i)--(iv) in Fig. 5.}
\label{tab3}
\begin{tabular}{lcccc}
\hline
   & (i) & (ii) & (iii) &(iv) \\
\hline
$\theta$  & $\pi/4$  & $\pi/4$  & $0.1$ & $0.1$  \\
$\m2$ (GeV) & $200$ & $300$ & $200$ & $300$ \\
\hline
\end{tabular}
\end{table}
\begin{figure}[htb]
\vspace{6cm}
\includegraphics{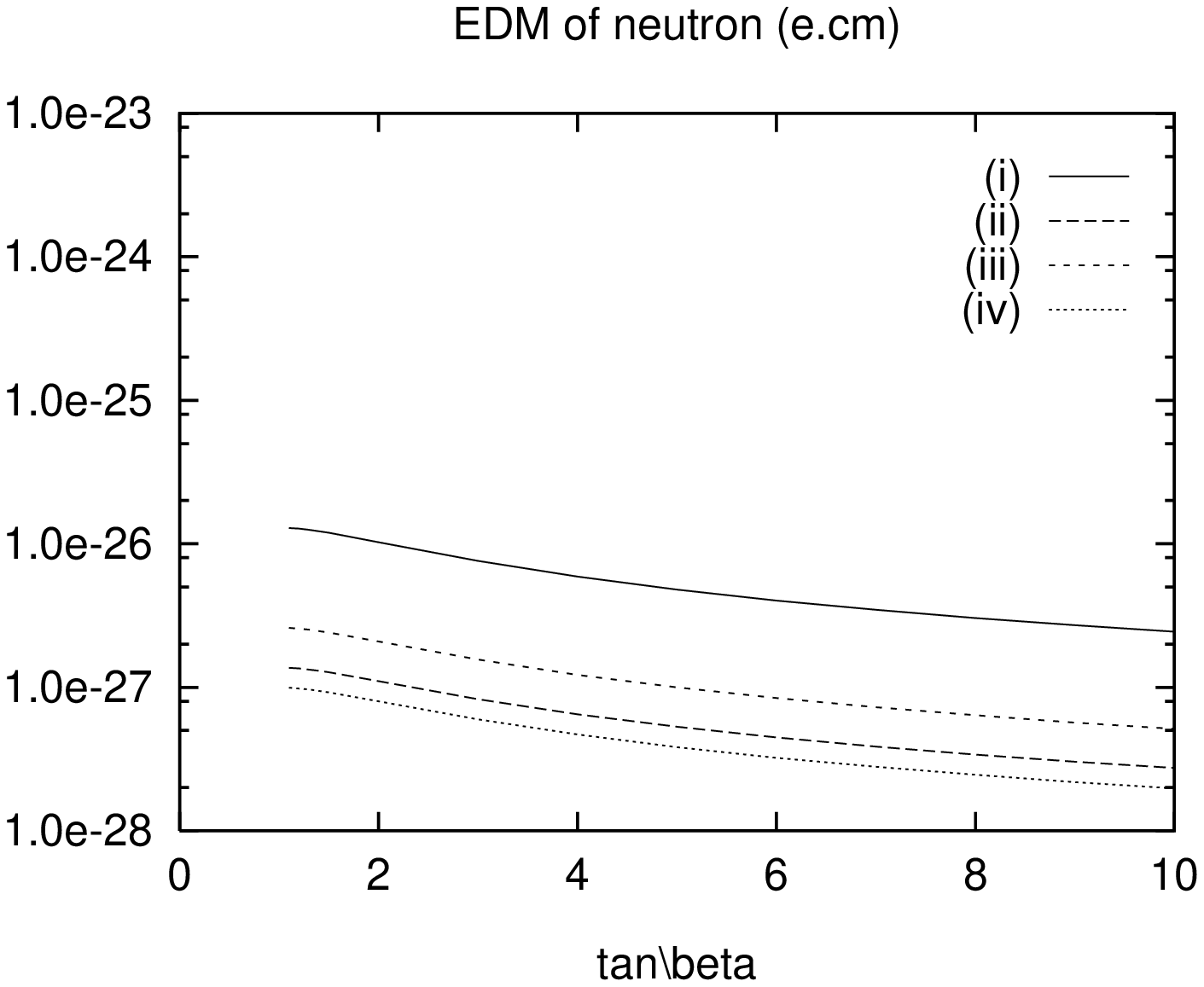}
\caption{The EDM of the neutron induced by the 
        $W$-boson EDM for $\theta=\pi/4$. 
        The values of $\m2$ and $|\mH|$ for curves (i)--(iv) 
         are listed in Table 2.}
\label{fig6}
\end{figure}         
     In Fig. \ref{fig5} we show the absolute value of 
the neutron EDM induced by the $W$-boson EDM for $\tb=2$ 
as a function of $|\mH|$. 
For $\theta$ and $\m2$ we have taken four sets of values given in 
Table \ref{tab3}.   
For $\theta\sim 1$, $\tb\sim 1$, and $\m2, |\mH|\sim 100$ GeV  
the magnitude of the neutron EDM is around 
$10^{-26} e$cm, which is smaller than the present 
experimental upper bounds by only one order of magnitude.  
Since the squark masses are at least 1 TeV for $\theta\gsim 0.1$,  
the magnitude of the neutron EDM arising from two-loop diagrams 
with the squarks becomes much smaller than $10^{-26} e$cm.   

     In Fig. \ref{fig6} the neutron EDM is plotted as a function of $\tb$ 
for four sets of values of $\m2$ and $|\mH|$ 
given in Table \ref{tab2}.  
The CP-violating phase is taken for $\theta=\pi/4$.  
The EDM of the neutron decreases as $\m2$ or $|\mH|$ increases.  
We can also see its clear dependence on $\tb$.  
As $\tb$ increases, the EDM decreases.  
This is contrary to the $\tb$ dependence of the 
neutron and electron EDMs induced by the one-loop 
diagrams as shown in Fig. \ref{fig3}.  

     The EDM of the electron induced by the $W$-boson EDM have a 
value smaller than the neutron EDM by one order of magnitude, 
similarly depending on the SSM parameters.  

\begin{figure}[htb]
\vspace{6cm}
\includegraphics{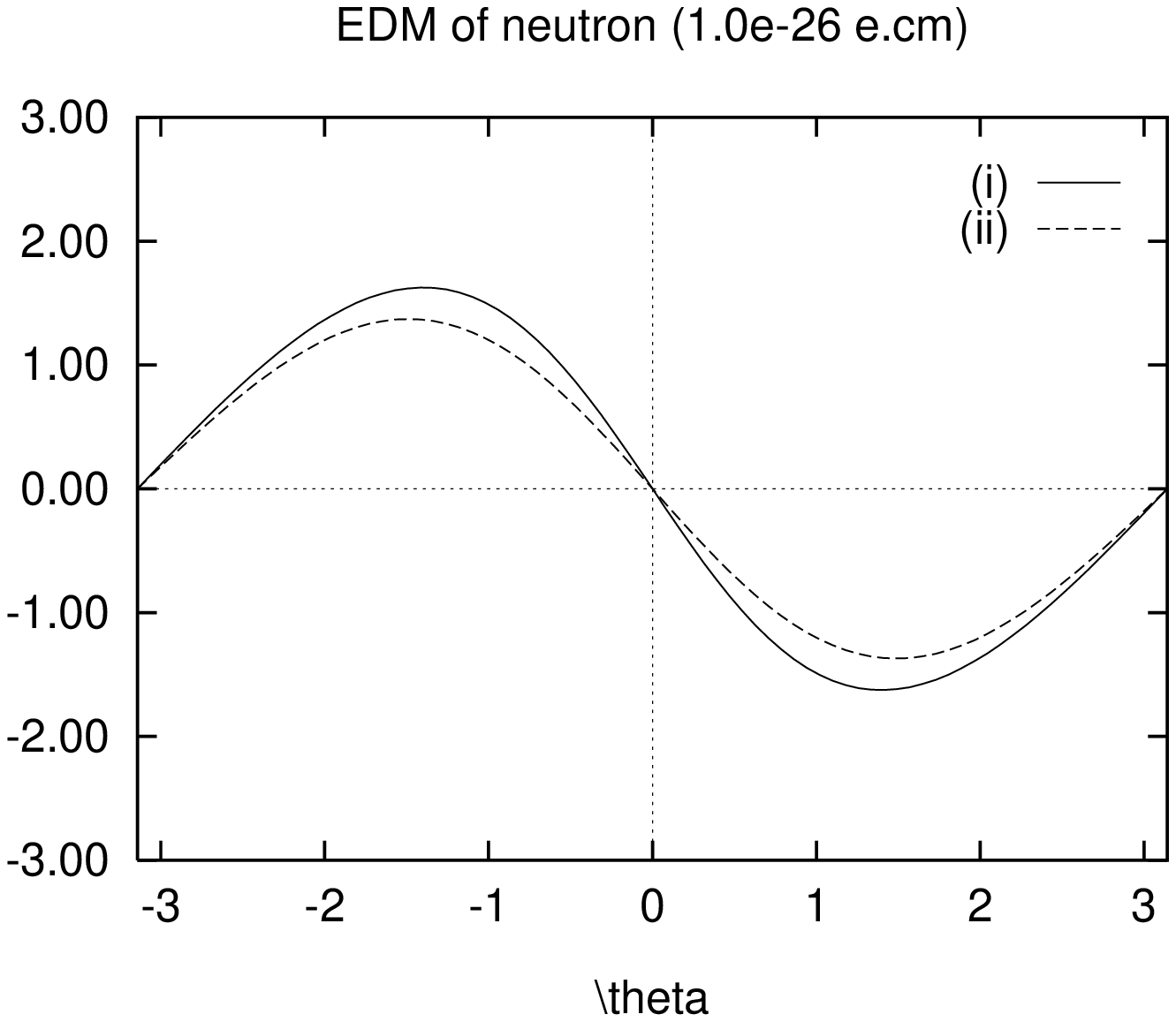}
\caption{The neutron EDM from the one-loop contributions 
      and that from the two-loop contributions,  
      which are represented respectively by curves (i) 
      and (ii).  
        The parameter values are $\m2=|\mH|=200$ 
        GeV and $\tb=2$.  For curve (i) the squark mass 
        is set for 10 TeV.}
\label{fig7}
\end{figure}         
     The quantitative difference between the one-loop contributions 
and the two-loop contributions to the EDM of  
the neutron or electron varies with the squark or slepton masses.  
For the masses of around 1 TeV, 
the one-loop contributions 
are larger than the two-loop contributions, whereas for the masses  
of around 10 TeV or larger, the latter  
can become larger than the former.  

     In Fig. \ref{fig7} we show the values of the neutron EDM 
at the one-loop and the two-loop levels as 
functions of the CP-violating phase $\theta$ for 
$\m2=|\mH|=200$ GeV and $\tb=2$.  
Curves (i) and (ii) respectively represent the EDM from 
the one-loop contributions 
and that from the two-loop contributions.   
The squark mass for the one-loop 
contributions has been taken for 10 TeV.  
The two types of contributions are comparable to each other.  
It is seen that 
the one-loop and the two-loop contributions have the same sign,  
which also holds for other reasonable values of the parameters 
and for the EDM of the electron.
Hence, for given values of  
the parameters for the gauge-Higgs sector,     
the magnitudes of the neutron and electron EDMs 
are expected to be larger than those obtained from 
the two-loop diagrams irrespectively of the squark 
and slepton masses.   

\section{Conclusion}

     We have discussed CP violation which 
originates from the gauge-Higgs sector in the SSM.   
If the baryon asymmetry of the universe is attributed 
to this source of CP violation, the CP-violating phase $\theta$ 
should be of order 0.1--1.   
Then, the EDMs of the neutron and electron,  
which are induced through 
the one-loop diagrams with the charginos 
and the squarks or sleptons,   
could be as large as their present experimental upper bounds.   
Accordingly, the parameters of the SSM are constrained.  
For $\theta\sim0.1$ the squarks and sleptons have to be 
heavier than or around 1 TeV, while the charginos and neutralinos 
could be of order of 100 GeV.  

     The EDM of the $W$ boson is induced at the one-loop 
level through the diagrams in which the charginos 
and neutralinos are exchanged.   
If $\theta$ is of order unity, 
the CP-odd form factor $\Fq$ in the effective Lagrangian 
for the $WW\gamma$ interactions could be of order of $10^{-4}$. 
Since this magnitude is far larger than the SM prediction,  
the $W$-boson EDM would be an interesting observable to search 
for supersymmetry.  
   
     The $W$-boson EDM could be examined 
through the EDMs of the neutron and electron.    
These EDMs receive contributions from the two-loop diagrams which 
contain the one-loop diagrams for the $W$-boson EDM.  
As a result, the large magnitude  
of the $W$-boson EDM implicates large magnitudes for the neutron and  
electron EDMs.  We have shown that  
the magnitudes of the neutron and electron EDMs at the two-loop 
level could be  
as large as $10^{-26}e$cm and $10^{-27}e$cm, respectively.    
These numerical outcomes are not so small compared to the experimental 
upper bounds at present. 
 
     The one-loop and two-loop contributions to the EDMs 
depend differently on the SSM parameters.  
For the squark and slepton masses of order of 1 TeV, 
the one-loop contributions are larger than the two-loop contributions.  
However, if those masses are around 10 TeV, the one-loop and the 
two-loop contributions could become comparable.  
Furthermore, these two types of   
contributions turned out to have the same sign.  
Therefore, the neutron and 
electron EDMs arising from the two-loop diagrams give theoretical lower 
bounds for given parameter values of the gauge-Higgs sector.   

     If the neutron and electron EDMs 
respectively have values of order of 
$10^{-26}e$cm and $10^{-27}e$cm, 
they will possibly be detected in the near future.  
Since the KM mechanism in the SM does not 
predict such large magnitudes  
for the EDMs,  
the detection of the EDMs 
could make the SSM a more credible candidate for the 
extension of the SM.   

\smallskip 
     The author thanks M. Aoki, T. Kadoyoshi, and A. Sugamoto for 
discussions.

\end{document}